\documentclass[12pt]{article}
\begin{document}
~\vskip 1truein
\begin{center}
{\bf \large
Wide-Field Astronomical Surveys in the Next Decade:\\
\smallskip
A White Paper for Astro2010}
\end{center}
\medskip
\begin{center}
\thispagestyle{empty}
{\it Michael A. Strauss\footnote{\rm strauss@astro.princeton.edu,
    1-609-258-3808} (Princeton),
J. Anthony Tyson (UC Davis),
Scott F. Anderson (Washington), 
T.S. Axelrod (LSST), 
Andrew C. Becker (Washington),
Steven J. Bickerton (Princeton),  
Michael R. Blanton (NYU), 
David L. Burke (SLAC), 
J.J. Condon (NRAO),
A.J. Connolly (Washington), 
Asantha Cooray (Irvine),
Kevin R. Covey (Harvard), 
Istv\'an Csabai (E\"{o}tv\"{o}s),
Henry C. Ferguson (STScI), 
\v Zeljko Ivezi\'c (Washington), 
Jeffrey Kantor (LSST), 
Stephen M. Kent (FNAL), 
G. R. Knapp (Princeton), 
Steven T. Myers (NRAO), 
Eric H. Neilsen, Jr. (FNAL), 
Robert C. Nichol (Portsmouth), 
M. Jordan Raddick (JHU), 
Baruch T. Soifer (IPAC), 
Matthias Steinmetz (Potsdam),
Christopher W. Stubbs (Harvard), 
John L. Tonry (Hawaii), 
Lucianne M. Walkowicz (Berkeley), 
R.L. White (STScI),
Sidney Wolff (NOAO), 
W. Michael Wood-Vasey (Pittsburgh),
Donald G. York (Chicago)
}
\end{center}

\begin{quote}
Wide-angle surveys have been an engine for new discoveries throughout
the modern history of astronomy, and have been among the most highly
cited and scientifically productive observing facilities in recent
years.  This trend is likely to continue over the next decade, as many
of the most important questions in astrophysics are best tackled with
massive surveys, often in synergy with each other and in tandem with
the more traditional observatories.  We argue that these surveys are
most productive and have the greatest impact when the data from the
surveys are made public in a timely manner.  The rise of the ``survey
astronomer'' is a substantial change in the demographics of our field;
one of the most important challenges of the next decade is to find
ways to recognize the intellectual contributions of those who work on
the infrastructure of surveys (hardware, software, survey planning and
operations, and databases/data distribution), and to make career paths
to allow them to thrive.
\end{quote}

\newpage
\setcounter{page}{1}
\section{Introduction}

  Very crudely speaking, breakthroughs in observational
  astronomy in the last fifty years have been driven by two types of
  facilities (often working in synergy): 
\begin{itemize} 
\item {\bf Observatories} are designed to allow detailed studies
  of individual objects or relatively small fields in a given
  waveband.  Much of the push towards telescopes of ever larger
  aperture is motivated by studies of individual objects.  
\item {\bf Survey facilities} are often dedicated telescopes (or
  systems of telescopes) with a wide field of view, which gather data
  on large numbers of objects, for use in a wide variety of scientific
  investigations.
\end{itemize}

Both types of facilities are driven by, among other things, new
technological developments in the field.  This has been especially
true for the opening of new wavelength regimes, and the history of
astronomy has taught us over and over again that there are
unanticipated surprises whenever we view the sky in a new way. 

Astronomers carry out the equivalent of experiments by discovering,  
and then studying, different sorts of astrophysical systems. Surveys
generate the list of available   
laboratories for such studies, and, as such, are central to progress  
in the discipline. Complete, unbiased surveys are the best technique  
we have both for discovering new and unexpected phenomena (Harwit
1981; Kellerman \& Sheets 1983), and for deriving the intrinsic  
properties of source classes so that their underlying physics can be  
deduced. 

Survey science tends to fall into several broad categories: 
\begin{itemize} 
\item Statistical astronomy, where large datasets of uniformly
  selected objects are used to determine distributions of various
  physical or observational characteristics.  Examples include
  measurements of the large-scale distribution of galaxies, or
  searches for stellar streams in the halo of the Milky Way.  Often,
  surveys are 
  designed as experiments to make very specific measurements along
  these lines, such as many of the CMB mapping surveys.  
\item Searches for rare and unanticipated objects.  Every major survey
  that has broken new ground in sensitivity, sky coverage or
  wavelength has made important serendipitous discoveries; surveys
  need to be designed to optimize the chances of finding the
  unexpected.  Examples include the discovery of pulsars,  of
  ultraluminous infrared galaxies by IRAS, and of supernova light
  echoes in the MACHO survey.  Some surveys are
  explicitly designed to look for very rare objects of a certain type,
  such as the planet searches by Corot and Kepler.
\item Surveys of the sky become a legacy archive for future
  generations, allowing astronomers interested in a given area of sky
  to ask what is already known about the objects there, to
  photometrically or astrometrically calibrate a field, or to select a
  sample of objects with some specific properties.  
\end{itemize}

In optical astronomy (the field in which most of the authors of this
white paper work), the state of the art for wide-field surveys for
many years was the
Palomar Observatory Sky Survey (POSS; 1948-1957), and its successors
in the 1980s on the UK Schmidt and at Palomar, which imaged the
entire celestial sphere with photographic plates.  This has been used
as a resource for a great deal of statistical astronomy: Abell's
famous cluster catalog and the major galaxy catalogs of the 1970s
such as the UGC come to mind, although its real power for quantitative
analysis came when it was digitized in the 1980s by STScI and other
teams.  However, the limitations of photographic film, especially in
sensitivity and linearity, meant that the next generation of surveys
had to wait until CCDs became large enough to be competitive with
film, telescope optics advanced to the point to allow wide-field focal
planes on large telescopes, and computers became powerful enough to
handle the resulting enormous data flow.

  The Sloan Digital Sky Survey (SDSS; 1998-present) was enabled by
  exactly these developments.  It has imaged roughly
  1/3 of the Celestial Sphere to $r \sim 22.5$ in five photometric
  bands, and obtained spectra of over 1.5 million quasars, galaxies, and
  stars.  The imaging camera, the largest astronomical camera in the
  world at the time it was built, has about 150 million pixels on the
  focal plane, and produces data at a rate of 5 Mbytes per second.  The
  CCDs themselves were among the most expensive components of the
  hardware, and the data rate was large enough that the survey would
  have been unthinkable with the computer power available a decade
  earlier. 

  The SDSS' core science goal was a three-dimensional map of the
  large-scale distribution of galaxies, but one of the lessons from
  this survey is that wide-field imaging and spectroscopy of the sky
  are fundamental for essentially all branches of observational
  astronomy.  The project has resulted in over 2200 refereed papers to
  date, the majority of which are authored by people outside the SDSS
  collaboration itself.  Indeed, the author lists of these papers
  include roughly 4000 unique individuals, an appreciable fraction of
  the world total of active research astronomers.  These papers cover
  a broad range of topics, from the structure of the asteroidal main
  belt, to the white dwarf luminosity function, to the structure of
  the Milky Way halo, to the dark matter masses of galaxies and the
  most distant quasars in the universe, to the large-scale structure
  studies for which the survey was designed.  The SDSS was the first
  or second most highly cited observatory facility in each of the
  years from 2003 to 2006 (Madrid \& Macchetto 2006, 2009), and is of
  comparable cost or cheaper than the other facilities with which it
  was compared, including HST and the 10-meter telescopes.

  The SDSS experience is not unique.  In the 1990s, the FIRST and NVSS
  20 cm surveys with the VLA, each covering
  thousands of square degrees, have become absolutely essential
  resources for the astronomical community.  The WMAP survey of the CMB
  sky has resulted in the most cited paper in the history of astronomy
  (Spergel et al. 2003).  The IRAS survey in the mid-infrared carried
  out in 1983 is still used by a large community of astronomers, and
  has resulted in over 5000 refereed papers, including the second-most
  cited paper in astronomy (Schlegel et al. 1998). The ROSAT survey of
  the sky at soft X-rays has resulted in over
  3500 papers since it was carried out a decade ago.  The Two-Micron
  All-Sky Survey has yielded 
  fundamentally new insights into the structure of the Milky Way and
  the coolest brown dwarfs.  There are of course many other examples.
  In each case, the data and resulting catalogs have been made
  available to the community in a scientifically useful form, and the
  majority of the papers produced by 
  them were written by astronomers outside the group of people
  responsible for producing the survey.  

  The exponential increase in survey data and the resulting science
  opportunities has resulted in the development of a new breed of
  scientist, the ``survey astronomer''.  These include both the people who do
  the very hard work of developing the infrastructure of these surveys
  (the ``builders''), and those who analyze these data for exciting
  science results (the ``miners'').  As we will argue below, one of
  the challenges of the next decade is to make sure that the work of
  this important new demographic component of our community is
  recognized, and that rewarding career paths are developed for them.

\section{The Next Decade}

The success of the SDSS and other surveys, and the important
scientific questions facing us today, have motivated astronomers to
plan the next generation of major surveys.  In the optical and
near-infrared in particular, wide-field cameras are being built for a
variety of telescopes.  On-going and planned post-SDSS imaging
surveys include the CFHT Legacy Survey, Pan-STARRS 1 and 4, the Dark
Energy Survey, SkyMapper, VISTA, Hyper-SuprimeCam on Subaru and the
Large Synoptic Survey Telescope.  These surveys will go appreciably
deeper than SDSS, and will open the time domain to study the variable
universe.  The next generation of spectroscopic surveys includes
SDSS-III, The Hobby-Eberly Telescope Dark 
Energy Experiment (HETDEX), LAMOST, RAVE, several of the concepts
associated with JDEM planning, and WFMOS, the
wide-field spectroscopic capability planned by the Gemini community to
be placed on the Subaru telescope.  There are similarly ambitious plans in
other wavebands; WISE will survey the entire sky from 3 to 20 microns,
eROSITA is a planned medium-energy X-ray sky survey, 
and the radio community is planning ASKAP as a pathfinder for the
Square Kilometer Array, which will carry out major redshift surveys of HI
in galaxies.  Note that some of these surveys use dedicated
telescopes, but the majority will work on already existing facilities,
and thus will observe only a fraction of the available clear nights.  


  These surveys are driven by some of the most important questions
in astrophysics today, as can be seen by the many science white papers
submitted to the Decadal Survey.  A partial list of the themes follows:
\begin{itemize} 
\item {\it Cosmological Models, and the Nature of Dark Energy and Dark
  Matter:}  Surveys of the Cosmic Microwave Background, the large-scale
  distribution of galaxies, the redshift-distance relation for
  supernovae, and other probes, have
  led us to the fascinating situation of having a precise cosmological
  model for the geometry and expansion history of the universe, whose
  principal components we simply do not understand.  A major challenge
  for the next decade will be to gain a physical understanding of dark
  energy and dark matter.  Doing this will require wide-field surveys
  of gravitational lensing, of the large-scale distribution of
  galaxies, and of supernovae, as well as next-generation surveys of
  the CMB (including polarization).  Many of the next generation of
  surveys will carry out aspects of this research program.  
\item {\it The Dark Ages:} One of the best probes we have of the
  physics of the universe between redshifts of 30 and 6 is emission
  and absorption of the 21 cm line of neutral hydrogen over large
  angular scales.  The technical
  challenges to mapping this (from both terrestrial and Galactic
  foregrounds) are formidable, but a number of surveys are being
  developed or planned, including the Mileura Widefield Array, LOFAR,
  the SKA, and 
  others.  
\item {\it The Evolution of Galaxies:} Surveys carried out with the
  current generation of 10-meter-class telescopes in synergy with deep
  X-ray (Chandra, XMM) and infrared (Spitzer) imaging have resulted in
  the outline of a picture of how galaxies evolve from redshift 7 to
  the present.  We now have a rough estimate, for example, of the star
  formation history of the universe, and we are starting to develop a
  picture of how the growth of supermassive black holes is coupled to,
  and influences, the growth of galaxy bulges.  But the development of
  galaxy morphologies and the dependence on environment are among many
  poorly understood questions, and the next generation of surveys
  promises to yield insights into these problems. 
\item {\it Structure of the Milky Way:} Encoded in the structure,
  chemical composition and kinematics of stars in our Milky Way is a
  history of its formation.  Surveys such as 2MASS and SDSS have
  demonstrated that the halo has grown by accretion and
  cannibalization of companion galaxies, and it is clear that the next
  steps require deep wide-field photometry, proper motions, and
  spectra to put together the story of how our galaxy formed.  This is
  one of the principal drivers for a variety of surveys, including
  SDSS-III, RAVE, GAIA, WFMOS, Pan-STARRS, SkyMapper, LSST, and
  LAMOST, among others.    
\item{\it The Variable Universe:} Variable and transient phenomena
  have historically led to fundamental insights into subjects ranging
  from the structure of stars to the most energetic explosions in the
  universe to cosmology (think SN Ia).  Existing surveys leave large
  amounts of discovery parameter space (in waveband, depth, and
  cadence) as yet unexplored, and the next generation of surveys is
  designed to start filling these gaps.
\item{\it Asteroids in the Solar System:} We now know of over 1000
  minor planets in orbits beyond that of Neptune, and are realizing that
  they fall into a wide variety of dynamical classes which encode
  clues to the formation of the solar system.  With over
  100,000 main-belt asteroids with known orbits, we can look for subtle
  correlations between their physical properties and 
  dynamics.  Many asteroids live on Earth-crossing orbits, and
  Congress has mandated that NASA catalog 90\% of all 
  potentially hazardous asteroids larger than 140 meters in diameter.  
  Dramatically increasing the available sample sizes are major
  goals of Pan-STARRS and LSST, among other surveys. 
\end{itemize}

Many of these science goals require similar data; in particular, 
wide-field repeated deep optical imaging can contribute to nearly all
of them.  The observing efficiency of a survey (\'etendue) scales with the
telescope size, field of view of the instrument, and duty cycle
(fraction of the time on the sky); with high enough \'etendue, a
single cadence gathers data that can be used by multiple science
projects (Ivezi\'c et al.~2008).   Many of these science goals need a
major increase in spectroscopic survey capability as well; 
for example, surveys like DEEP2 and VVDS only whet the appetite for
spectroscopic survey volumes comparable to the SDSS main galaxy sample
in several redshift bins to high redshift.  

It is also worth mentioning that both the hardware and computational
technical challenges, and the exciting science opportunities, are
attracting scientists from other disciplines, including high-energy
physics, statistics, and computer science.  This is a wonderful
opportunity for astronomers to learn from, and take advantages of the
advances in, these other fields, and to grow the community of
scientists interested in survey science.

\section{Lessons Learned, and Recommendations}

What have the current generation of surveys, and the planning for the
next generation of surveys, taught us?  

\subsection*{Data quality is paramount}
Large projects are always starved for time and money, and an obvious
temptation is to cut corners by skimping on data quality: not
ferreting out all the systematic errors in photometry, astrometry or
wavelength calibration, not optimizing algorithms to reduce false
positives in the data stream, or not putting sufficient effort into
quality assurance tools to catch problems as they crop up in the data.
A lesson that the SDSS team found itself learning over and over, and
is shared by other surveys, is that it is always cheaper to do things
right the first time.  Survey requirements are defined by first deciding on
the core science goals, then designing a telescope, instrument and
survey strategy that will meet these goals.  One then asks what the
data quality that these, together with the laws of physics allow, and
designs rigorous quality validation tools to verify that the data meet
these requirements.  Doing this will give a much more uniform dataset
and enable science well beyond that anticipated at the time the survey
was designed.

  Of course, such an approach will inevitably result in tensions
  between the need to keep on budget and schedule, and the desire to
  do things right.  There are no simple answers to balancing these
  two, and it means that a survey must budget with realistic
  contingencies to allow flexibility when problems inevitably arise.  

\subsection*{Software is important}
It is a truism that the software necessary to run a major survey is at
least as large an intellectual effort, and requires similar
resources, as
the design and building of the telescope and instruments.  The
software must therefore be included in plans and budgets from the
beginning of any substantial survey, especially those surveys which
break new ground in terms of the quantity or nature of the data they
gather.  Moreover, writing the software is not a one-off deal; the
survey will continue to need further pipeline and quality assurance
work throughout its lifetime (the final data release paper of SDSS-II
(Abazajian et al.~2008), coming out a decade after first light,
included 18 pages describing the pipeline improvements made in the
final year of operations).  

\subsection*{Support the people who make the surveys happen}
Major surveys involve large groups of talented people working on the
instrument, survey design, software pipelines, observations,
databases, and other aspects.  These people are often at early stages
of their careers: graduate students, postdocs, or assistant
professors.  If they are working on the early stages of a many-year
survey, they will not necessarily have the data or the time needed to
write first-author papers.  Lead-author papers have traditionally been
the currency by which astronomers are traditionally recognized, and
those who work on the survey infrastructure are often at a
disadvantage in career advancement.   A major challenge in the
next decade will be finding ways to change the astronomical culture to
more directly recognize the tremendous intellectual contribution of
those people working on survey infrastructure, and to understand that
papers are not the only mark of productivity and creativity in the
field.  This can be done both in traditional academia, for promotion
to faculty positions, and also through non-academic environments in
which people working on survey infrastructure may be supported.  As an
example of the latter, consider IPAC, which has played a major role in
IRAS and 2MASS, among other surveys.  

It is worth recognizing that an increasing number of astronomers are
building successful careers by carrying out science enabled by
surveys.  It was not too long ago that ``armchair astronomer'' was a
derogatory term meant for those working with data that they themselves
didn't obtain at the telescope, but a glance at the current generation
of assistant professors around the country reveals many who have made
ground-breaking discoveries using data drawn from large surveys.  The
best of those people have managed to work on both science and
infrastructure of the surveys; working ``in the trenches'' is the best
way to understand the data in all its 
nuances, and therefore be able to exploit it for all its scientific
value.

The National Science Foundation has been very generous in its support
of surveys.  The typical model has been to support the construction
and operation of a given survey, leaving the scientific analysis of
the data (i.e., the process that results in papers) to be funded
separately.  We are of two minds about this.  On the one hand, surveys
are usually designed with very specific scientific goals, and cannot
claim to succeed until those goals are met; in this context, funding
the analysis to the point of completed papers makes sense for many
projects.  NASA tends to operate this way, for both surveys and
observational facilities; the grant that enabled the construction of
WMAP and its resulting data products also funds the core scientific
papers the team writes.  Giving the young people working on the
infrastructure of the project some science support can be an important
boost to their careers.  We have seen cases in which survey builders 
apply for grants to do the science that their surveys that they have
spent the previous decade bringing to fruition were
designed to do, only to be turned down with referees remarking on their
apparent lack of published papers in the previous five years!  With
this in mind, it might make sense to set aside funds
specifically for survey builders to reap the scientific benefits of the 
data that they helped create.  

 On the other hand, as we argue below, most of the
good science for any given survey will ultimately be done by people outside the
collaboration (i.e., the ``miners''), and the existing grants
mechanism within the NSF has worked adequately to support those people
(given the limitations of the huge oversubscription ratio that the NSF
grants program faces).   

\subsection*{The data must be made public}
We have already emphasized that major surveys have scientific value
far beyond that for which the surveys were originally designed.  This
means, in particular, that the pipelines and databases should be
developed with the general user in mind, not just those working on the
core scientific goals of the project.  Moreover, this means that a
survey will enable far more science than the builders of the
survey will be able to carry out themselves.  Therefore a survey must
plan to make its data and software pipelines public and properly
documented in a form that allows the full scientific community to use them.  A
proprietary period, whereby those who built the project get exclusive
access to the data, may be considered necessary in the beginning as a
motivation for people to put in the work; big projects will also
need time to analyze their data and check its quality.  But it is hard
to imagine circumstances in which this proprietary period should be
longer than a year or two.  And of 
course, proprietary periods make no sense for synoptic surveys, where
rapid follow-up on a wide variety of other facilities is key.  Indeed,
the trend in the field is towards ``open source/open data'', that is,
surveys with no proprietary period for the data, and with
publicly available software (a welcome trend for all astronomy,
not just surveys!).  Despite the lack of exclusive privileged
access to the data, 
those people working on the survey itself have an inside track; their
intimate knowledge of the data and survey characteristics with 
all its quirks offers a significant advantage in getting interesting results.  

Major surveys are increasingly a resource for more than the broad
astronomical community: they are gathering growing interest from
the general public, from school children to interested amateurs.  Data
from surveys are being incorporated into K-12 classroom
activities, and websites like Google Sky and Microsoft's WorldWide
Telescope have reached millions.  The
overwhelming success of the Galaxy Zoo project\footnote{\tt
  http://www.galaxyzoo.org}, which has involved over 150,000 members
of the public to gain real scientific insights into the nature of
galaxies, tells us that the public and the astronomical community
can gain both from active outreach, and from exploring creative ways to make
public databases accessible to non-professionals.  

Needless to say, distributing data publicly does not come for free;
as we now argue, this requires sophisticated databases and extensive
documentation, which must be budgeted for when the survey is first
designed. 

\subsection*{Data distribution and archiving should be built in from
  the beginning} 
We live in the era of large databases.  The surveys of the current
decade have data sizes measured in terabytes, and those of the next
decade will exceed petabytes.  This is orders of magnitude more data
than can usefully be examined as flat files, or which can easily be
distributed by putting it online for people to download.
Databases for distributing and examining the data must be built
into the survey plans from the beginning, and must be designed for
the sort of scientific analyses that people will do; the
database has to be general-purpose enough to allow for scientific
projects that were unanticipated by the survey designers at the
beginning of the project.  This is perhaps an obvious
statement for the purposes of making the data public, but it is just as
crucial for distributing the data {\em within} a collaboration.
Modern surveys are often carried out with consortia spanning the
globe, and the researchers at the various member institutions will
need ways to access the data as early as possible.   Indeed, given the
sizes of the next generation of surveys, it becomes impractical to pull all
the data one needs for many scientific analyses to one's home
institution; surveys have to plan to provide
computational power along with the data themselves to the end
scientific user.  

This remains an issue long after the survey is completed; the survey
data will have archival value for decades.  Astronomers will 
want to use the astrometry to measure proper motions with very long
time baselines and to look for variable phenomena of all sorts,
and of course they will continue to mine the data for various scientific
projects\footnote{A wonderful example of the need for very
  long-term archives may be found in the title of a presentation at
  the last AAS meeting, {\it Front-line Recurrent Nova Science
    Requires Century Old Data} (Schaefer 2009).}.  Given the rapidity
with which computer and data storage systems change and become
obsolete (Rothenberg 1995), long-term data archiving becomes a real
challenge, and one that requires continuous attention.

\subsection*{Real project management is important}
Modern surveys are big, expensive projects involving large numbers of
astronomers who are typically spread between a number of institutions.
Such projects are much too large to be managed by the astronomers who
lead the projects scientifically, and they need professional project
management to keep track of budgets, schedules, and the responsibilities
of the different institutions.  There will be inevitable tensions
between the project managers and the scientists who are ensuring the quality
of the data, for which there are no easy answers.  
Keeping
communication lines open and maintaining mutual respect between distant
collaborators is a continuous challenge, best met with archived e-mail
exploders, frequent phone conferences, and clear lines of authority
and statements about requirements and responsibilities.  Face-to-face
meetings are essential, and should be held at least twice a year. 

\subsection*{Synergy between surveys, and with observatories}
The intercomparison of surveys allows science that would be impossible with
any one survey alone.  This comparison can be temporal (e.g.,
comparing the proper motion of an object between the POSS and the
SDSS; Munn et al. 2004) or across wavelength regimes (e.g., looking
for long-term optical counterparts to gamma-ray 
bursts).  The standards of the Virtual Observatory give us a mechanism
to make cross-survey comparisons easy, and most planned surveys
intend to follow these standards.  Indeed, many of the most important
astronomical problems we face require multiple probes via interlocking
surveys.  A major theme of the next generation of CMB mapping surveys, for
example, will be cross-correlating with surveys in the X-ray,
ultraviolet, and optical to ameliorate foreground contamination, and
to measure the Integrated Sachs-Wolfe Effect, the Sunyaev-Zel'dovich
effect, and the gravitational lensing of the CMB by foreground
structure.  

The discoveries made in surveys are often best exploited by detailed
study with other telescopes.  Unusual objects from an imaging survey
will require follow-up spectroscopy to determine their physical
nature; one can imagine, for example, a great deal of synergy of this
sort between the LSST and the GSMT.  Similarly, transient objects
such as the gamma-ray bursts which synoptic surveys will find, require
multi-wavelength follow-up over an extended period of time,
to allow these discoveries to be placed in astrophysical context.
We are particularly excited about plans for arrays of
robotically controlled moderate telescopes, designed specifically for
following up transients found in the next generation of synoptic
surveys.  

\section{Concluding Remarks}
Based on our experience with the surveys of the 1980s through the
present, the major astronomical surveys of the next decade have the
potential to revolutionize our understanding in many areas of modern
astrophysics.  In doing so, they will involve an appreciable fraction of the
worldwide astronomical community, at institutions from small
liberal-arts colleges to major research universities.  The
non-proprietary nature of data from these surveys means that students
and faculty are not limited by a lack of access to large observing
facilities in order to carry out meaningful and cutting-edge
research.  Thus the current and next generation of large surveys are
serving as a democratizing force in astronomy, helping to level the
playing field for researchers and students at smaller and less
well-endowed institutions.  Moreover, there
are tremendous opportunities to involve the general public in surveys, both
in educational activities and in real scientific enterprises such as
Galaxy Zoo.  

Survey astronomy will play a major role in the direction and
development of astronomical research in the next decade.  But surveys
are not easy: doing them well requires tremendous attention to data
quality, and a substantial allocation of resources for software,
database/data distribution systems, documentation, and long-term
archiving.  Carrying them off requires input from astronomers with a
very large range of skills, including survey, telescope, and
instrument design, software, databases, and scientific analysis.  As
surveys become ever more a part of the astronomical landscape, our
community has to find ways to support the growing community of
scientists who make them happen, and to fund the scientific
exploitation of these data.  There is a real demographic shift in the
community, with more and more scientists falling under the rubric of
``survey astronomer'', both the builders and the miners.  This is a
trend we should welcome and nurture, while making sure that those data
miners are fully experienced in, and cognizant of, the inner workings
of the surveys that they use, and that survey builders are given the
scientific support they need to exploit the surveys they help create.

\bigskip
\noindent $\bullet$ Abazajian, K. et al.~2008, arXiv:0812.0649\\
$\bullet$ Harwit, M. 1981, Cosmic Discovery (Sussex: Basic Books) \\
$\bullet$ Ivezi\'c, \v Z. et al. 2008, arXiv:0805.2366\\
$\bullet$ Kellerman, K., \& Sheets, J. 1983, Serendipitous Discoveries
in Radio Astronomy (Green Bank: NRAO)\\
$\bullet$ Madrid, J.P. \& Macchetto, D. 2006, BAAS, 38, 1286\\
$\bullet$ Madrid, J.P. \& Macchetto, D. 2009, arXiv:0901.4552\\
$\bullet$ Munn, J.A. et al. 2004, AJ, 127, 3034\\
$\bullet$ Rothenberg, J. 1995, Scientific American, January issue, page 42\\
$\bullet$ Schaefer, B.E. 2009, BAAS, \#213, \#320.05\\
$\bullet$ Schlegel, D.J, Finkbeiner, D.P., \& Davis, M. 1998, ApJ,
500, 525\\
$\bullet$ Spergel, D.N. et al.~2003, ApJS, 148, 175
\end{document}